\documentclass[10pt,journal,epsfig]{IEEEtran}

\usepackage[dvips]{graphicx}
\usepackage{graphicx}
\usepackage{amssymb}
\usepackage{cite}
\usepackage{subfigure}
\usepackage{amsmath}
\usepackage{amsthm}
\usepackage{setspace}
\usepackage{footnote}
\usepackage{enumerate}

\IEEEoverridecommandlockouts

\begin{document}

\title{Perturbation-Assisted PAPR Reduction for Large-Scale MIMO-OFDM Systems via ADMM}

\author{Hengyao Bao, Jun Fang,~\IEEEmembership{Member,~IEEE},
Zhi Chen,~\IEEEmembership{Member,~IEEE} and Tao
Jiang,~\IEEEmembership{Senior Member,~IEEE}
\thanks{Hengyao Bao, Jun Fang, and Zhi Chen are with the National Key Laboratory
of Science and Technology on Communications, University of
Electronic Science and Technology of China, Chengdu 611731, China,
Email: JunFang@uestc.edu.cn, chenzhi@uestc.edu.cn}
\thanks{Tao Jiang is with the School of Electronics Information and
Communications, Huazhong University of Science and Technology,
Wuhan 430074, China, Email: Tao.Jiang@ieee.org}
\thanks{This work was supported in part by the National Science
Foundation of China under Grant 61522104.}}

\maketitle


\begin{abstract}
We consider the problem of peak-to-average power ratio (PAPR)
reduction for orthogonal frequency-division multiplexing (OFDM)
based large-scale multiple-input multiple-output (MIMO) systems. A
novel perturbation-assisted scheme is developed to reduce the
PAPRs of the transmitted signals by exploiting the redundant
degrees-of-freedom (DoFs) inherent in the large-scale antenna
array. Specifically, we introduce artificial perturbation signals
to the frequency-domain precoded signals, with the aim of reducing
the PAPRs of their time-domain counterpart signals. Meanwhile, the
additive perturbation signal associated with each tone is
constrained to lie in the null-space of its associated channel
matrix, such that it does not cause any multi-user inference or
out-of-band radiations. Such a problem is formulated as a convex
optimization problem, and an efficient algorithm is developed by
resorting to the variable splitting and alterative direction
method of multipliers (ADMM) techniques. Simulation results show
that the proposed method has a fast convergence rate and achieves
substantial PAPR reduction within only tens of iterations. In
addition, unlike other precoding-based PAPR reduction methods, our
proposed method which introduces perturbation signals to the
precoded signals is independent of the precoding stage and thus
could be more suitable for practical systems.
\end{abstract}






\section{Introduction}
Large-scale multiple-input multiple-output (MIMO), also known as
very-large or massive MIMO, is a very promising technology for the
next generation wireless communication systems. In large-scale
MIMO systems, a large number of antennas are equipped at the base
station (BS), simultaneously serving a much smaller number of
users sharing the same time-frequency resource. In addition to a
higher throughput, large-scale MIMO systems have the potential to
improve the energy efficiency and enable the use of inexpensive
low-power components. These advantages render large-scale MIMO an
appealing technology for future wireless communication systems.

Due to the delay spread of wireless channels, broadband wireless
communications generally suffer from frequency-selective fading.
The most widely used technique to deal with the
frequency-selective fading is the orthogonal frequency division
multiplexing (OFDM), in which digital symbols are independently
encoded on multiple ``orthogonal" sub-carriers. MIMO-OFDM has been
adopted as a standard air interface technique in many real
wireless communication systems, such as LTE-A \cite{LTE-A}, WiMAX
\cite{WiMAX} and Wi-Fi \cite{Wi-Fi}. However, OFDM-modulated
signals usually incur a high peak-to-average power ratio (PAPR),
due to the fact that phases of sub-carriers are independent of
each other and may combine in a constructive or destructive
manner. To avoid signal distortions and out-of-band radiations,
high-resolution digital-to-analog converters (DACs) and linear
power amplifiers need to be used for each antenna, which is not
only expensive but also power-inefficient. In particular, the cost
becomes unaffordable when the number of antennas is large, which
makes large-scale MIMO systems impractical. Therefore, it is of
crucial importance to reduce the PAPR of massive MIMO-OFDM systems
to facilitate low-cost and power-efficient hardware
implementations.

Over the past years, a plethora of PAPR reduction techniques have
been proposed for single-input single-output (SISO) systems
\cite{Clipping,TR,ACE,SLM,PTS,JiangYang05} and point-to-point MIMO
systems \cite{HanLee05,FischerHoch06,JiangWu08,TsiJones10}. The
extension of these schemes to the multi-user (MU) MIMO systems,
however, is not straightforward because joint signal processing at
the receiver side is impossible as users are spatially
distributed. In \cite{PrabhuEdfors14}, a PAPR reduction scheme
similar to the tone reservation (TR) \cite{TR} was developed for
large-scale MU-MIMO-OFDM systems, where the amplitude clipping is
used for some transmit antennas to reduce the PAPR, while other
antennas are reserved to compensate for the distortions caused by
the clipping. This method \cite{PrabhuEdfors14} has a low
computational complexity. But those antennas reserved for
compensation may incur large PAPRs. A precoding-based PAPR
reduction scheme was proposed in \cite{StuderLarsson13} for
large-scale MIMO-OFDM systems. The proposed method, through
designing a suitable precoding matrix, aims to reduce the PAPR of
the transmitted signal and, meanwhile, remove the multiuser
interference (MUI). Specifically, the joint PAPR reduction and MUI
cancelation problem was formulated as a linear constrained
$\ell_\infty$ optimization and a fast iterative truncation
algorithm (FITRA) was developed. Following \cite{StuderLarsson13},
efficient approximate message passing (AMP)-based Bayesian methods
\cite{BaoFang2016,ChenWang15} were also developed for joint PAPR
reduction and MUI cancelation for large-scale MIMO-OFDM systems,
in which the problem was formulated as searching for a low PAPR
solution to an underdetermined linear system.




In this paper, we develop a novel perturbation-assisted approach
to address the PAPR reduction problem for large-scale MIMO-OFDM
systems. Our proposed method introduces artificial perturbation
signals to the frequency-domain precoded signals. The perturbation
signals are devised to reduce the PAPRs of the time-domain
counterpart signals. Meanwhile, the additive perturbation signal
associated with each tone is constrained to lie in the null-space
of its associated channel matrix. This null-space constraint
guarantees that the additive perturbation signals cause no
multi-user interference or out-of-band radiations. The design of
the perturbation signals can be formulated as a constrained convex
optimization problem. By resorting to the variable splitting and
alterative direction method of multipliers (ADMM) techniques, we
develop an efficient algorithm to solve the optimization problem.
Compared with existing methods, e.g.
\cite{PrabhuEdfors14,StuderLarsson13,BaoFang2016,ChenWang15}, our
proposed method has the following advantages:
\begin{itemize}
\item  Most PAPR reduction schemes for large-scale MIMO-OFDM systems are precoding-based, e.g.
\cite{StuderLarsson13,BaoFang2016,ChenWang15}. These methods
reduce the PAPR through designing the precoding matrices or
devising the precoded signals directly. Nevertheless, in some
practical systems, e.g. in LTE-A systems, the precoding matrices
are usually chosen from a fixed codebook. Hence those
precoding-based schemes may be impractical for real systems. In
contrast, our proposed approach is independent of the precoding
design, and thus is free of this issue and compatible with real
systems.
\item Our proposed perturbation-assisted method does not cause
any additional multi-user interference (MUI) and out-of-band
radiations. If a zero-forcing precoding scheme is employed, then
perfect MUI cancelation can be achieved by our proposed method,
whereas other precoding-based methods (e.g.
\cite{StuderLarsson13,BaoFang2016,ChenWang15}) cannot guarantee
complete MUI cancelation. For example, the FITRA algorithm
\cite{StuderLarsson13} needs to choose an appropriate
regularization parameter to ensure a small amount of MUI.
\item  The proposed method has a low
computational complexity. Besides, numerical results show that the
proposed scheme has a fast convergence rate and achieves a
substantial PAPR reduction within tens of iterations, which are
amiable merits for practical systems.
\end{itemize}



The reminder of this paper is organized as follows. In Section II,
we discuss the system model and the PAPR reduction problem. A
perturbation-assisted PAPR reduction scheme is proposed in Section
III, where the PAPR reduction is formulated as a constrained
convex optimization problem. An efficient ADMM-based algorithm is
developed in Section IV to solve the optimization problem.
Simulation results are provided in Section V, followed by
concluding remarks in Section VI.

\emph{Notations:} Bold lowercase letters (e.g. $\boldsymbol{x}$)
denote column vectors, bold lowercase letters with a superscript
$(\cdot)^r$ (e.g. $\boldsymbol{x}^r$) denote row vectors, and bold
uppercase letters (e.g. $\boldsymbol{X}$) denote matrices. For a
$M\times N$-dimensional matrix $\boldsymbol{X}=\{x_{mn}\}$, we use
$\boldsymbol{x}_m$ to designate the $m$th column, and
$\boldsymbol{x}^r_n$ to designate the $n$th row. The superscripts
$(\cdot)^{*}$, $(\cdot)^{T}$ and $(\cdot)^{H}$ represent the
conjugate, transpose and conjugate transpose, respectively. In
addition, we use $\|\boldsymbol{x}\|_{2}$ and
$\|\boldsymbol{x}\|_{\infty}$ to denote the $\ell_{2}$-norm and
$\ell_{\infty}$-norm of vector $\boldsymbol{x}$, respectively, and
use  $\|\boldsymbol{X}\|_{F}$ to stand for the Frobenius norm of
matrix $\boldsymbol{X}$. The $N\times N$ identity matrix and the
$M\times N$ zero matrix are denoted by $\boldsymbol{I}_{N}$ and
$\boldsymbol{0}_{M\times N}$, respectively.


\section{Preliminaries}
\subsection{System Model}
\begin{figure*}[!t]
\centering
\includegraphics[width=17.9cm]{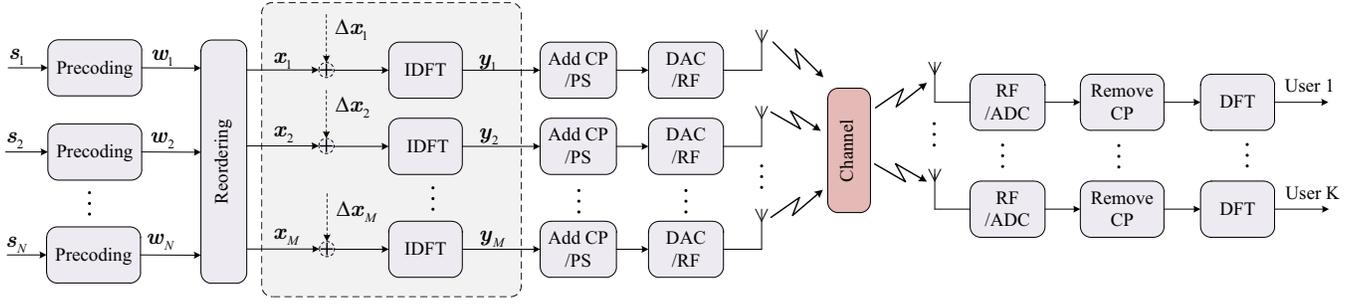}
\caption{System model of the large-scale multi-user MIMO-OFDM
downlink scenario, with $N$ OFDM tones, $\!M$ transmit antennas
and $K$ independent single-antenna users. Perturbation signals
$\{\boldsymbol{\Delta x}_m\}$ are added to the frequency domain
signals $\{\boldsymbol{x}_m\}$ to reduce the PAPR.}
\label{systemmodel}
\end{figure*}

The system model of the large-scale multi-user MIMO-OFDM downlink
scenario is depicted in Fig. \ref{systemmodel}. We assume that the
BS, equipped with $M$ transmit antennas, simultaneously serves $K$
single-antenna users, where $K\ll M$. The number of OFDM tones
(sub-carriers) is assumed to be $N$, and the signal vector
$\boldsymbol{s}_{n}\in\mathbb{C}^{K\times1}$ comprises the
modulated symbols associated with the $n$-th tone for the $K$
users. To shape the spectrum of the transmit signals, OFDM tones
are usually divided into two complementary sets $\mathcal{T}$ and
$\mathcal{T}^\mathcal{C}$, where the tones in set $\mathcal{T}$
are used for data transmission, and the tones in set
$\mathcal{T}^\mathcal{C}$ are used for guard bands which are
located at both ends of the spectrum. Moreover, for each tone
$n\in\mathcal{T}$, each symbol of $\boldsymbol{s}_{n}$ is chosen
from a complex-valued signal alphabet $\mathcal{B}$. For each tone
$n\in\mathcal{T}^\mathcal{C}$, we set
$\boldsymbol{s}_{n}=\boldsymbol{0}_{K\times1}$ such that no signal
is transmitted on the guard band.

For large-scale multi-user MIMO-OFDM systems, precoding needs to
be performed at the BS to eliminate multi-user interference (MUI)
at the receivers. The signal vector $\boldsymbol{s}_n$ can be
linearly precoded as
\begin{align}
\boldsymbol{w}_n=\boldsymbol{P}_n\boldsymbol{s}_n
\label{precoding}
\end{align}
where $\boldsymbol{w}_n\in\mathbb{C}^{M\times1}$ is an
$M$-dimensional precoded vector with its $m$th entry transmitted
through the $m$th antenna over the $n$th sub-carrier, and
$\boldsymbol{P}_n\in\mathbb{C}^{M\times K}$ corresponds to the
precoding matrix for the $n$th tone. Zero-forcing (ZF) is a
precoding scheme that aims at eliminating the MUI completely.
Since $K\ll M$, there are an infinite number of precoding matrices
that can achieve perfect MUI cancelation, among which the most
widely used form is
\begin{align}
\boldsymbol{P}_n^{\text{ZF}}=\boldsymbol{H}_n^H(\boldsymbol{H}_n\boldsymbol{H}_n^H)^{-1}
\label{ZF}
\end{align}
where $\boldsymbol{H}_{n}\in\mathbb{C}^{K\times M}$ is the channel
matrix associated with the $n$th tone. In this paper, we assume
that the channel matrices $\{\boldsymbol{H}_{n}\}$ are perfectly
known at the BS. Besides the ZF, other widely used precoding
schemes include matched filter (MF) precoding and minimum
mean-square error (MMSE) precoding \cite{Fischer2002}.

After precoding, all precoded signals $\boldsymbol{w}_{n}, \forall
n$ are reordered to $M$ transmit antennas for OFDM modulation,
\begin{align}
\boldsymbol{X}\triangleq[\boldsymbol{x}_{1}\cdot\cdot\cdot\boldsymbol{x}_{M}]=
[\boldsymbol{w}_{1}\cdot\cdot\cdot\boldsymbol{w}_{N}]^{T},
\label{reorder}
\end{align}
where $\boldsymbol{x}_{m}\in\mathbb{C}^{N}$ represents the
frequency-domain signal to be transmitted from the $m$th antenna.
In practice, a normalization may be applied to the
frequency-domain signal $\boldsymbol{X}$ to ensure unit or fixed
transmit power. This normalization is omitted here for simplicity.
The time-domain signal
$\boldsymbol{Y}\triangleq[\boldsymbol{y}_{1}\cdot\cdot\cdot\boldsymbol{y}_{M}]$
is obtained by performing an inverse discrete Fourier transform
(IDFT) of $\{\boldsymbol{x}_{m}\}$. To avoid the intersymbol
interference (ISI), a cyclic prefix (CP) is added to the
time-domain samples at each antenna. Finally, these samples are
converted to analog signals and transmitted.

At the receivers, after removing the CPs, the frequency-domain
signal can be obtained by performing a discrete Fourier transform
(DFT). Specifically, the signal received by $K$ users can be
expressed as
\begin{align}
\boldsymbol{r}_{n}=\boldsymbol{H}_{n}\boldsymbol{w}_{n}+\boldsymbol{e}_{n},
\quad \forall n\label{transmition}
\end{align}
where $\boldsymbol{r}_{n}\in\mathbb{C}^{K\times1}$ is the received
frequency-domain signal associated with the $n$th tone, and
$\boldsymbol{e}_{n}\in\mathbb{C}^{K\times1}$ denotes the receiver
noise whose entries obey i.i.d circularly-symmetric complex
Gaussian distribution with zero-mean and variance $N_0$. If ZF
precoding is employed, the MUI can be perfectly removed as we have
$\boldsymbol{r}_{n}=\boldsymbol{s}_{n}+\boldsymbol{e}_{n}, \forall
n$.

\subsection{Peak-to-Average Power Ratio (PAPR)}
OFDM is a digital multi-carrier modulation scheme which encodes
digital symbols on multiple ``orthogonal'' sub-carriers.
Specifically, given a frequency-domain signal $\boldsymbol{x}_m$
at the $m$th transmit antenna, the corresponding continuous time
domain OFDM signal can be written as
\begin{align}
y_m(t)=\frac{1}{\sqrt{N}}\sum_{n=0}^{N-1}x_{mn}\cdot e^{j2\pi n\triangle ft},~ 0\leq t<T
\label{continuous-ofdm}
\end{align}
where $T$ denotes the symbol duration, and $\triangle f=1/T$ is
the sub-carrier spacing that is carefully devised to make
sub-carriers orthogonal to each other. As can be seen from
(\ref{continuous-ofdm}), these sub-carriers may combine in a
constructive or a destructive manner since the phases of
sub-carriers are independent of each other. Therefore,
OFDM-modulated signals typically exhibit a large dynamic range,
which can be characterized by the peak-to-average power ratio
(PAPR) metric. The PAPR of a signal is defined as the ratio of the
peak power of the signal to its average power, that is
\begin{align}
\textsf{PAPR}\left(y_m(t)\right)=\frac{\max\limits_{0\leq{t}\leq{T}}|y_m(t)|^2}
{1/T\cdot\int_0^{T}|y_m(t)|^2dt}
\end{align}
Signals with a large dynamic range are usually susceptible to
non-linear RF components. To avoid undesirable out-of-band
radiation and in-band distortions, high-resolution DACs and linear
power amplifiers are required at the transmitter. Nevertheless,
these linear components are not only expensive but also
power-inefficient. Thus, PAPR reduction is crucial to facilitate
low-cost and power-efficient hardware implementations for
large-scale MIMO-OFDM systems.



Clearly, the PAPR of the analog signal $y_m(t)$ is of our concern.
Nevertheless, analog signals are not amiable for calculation and
analysis. To address this issue, we consider a sampled version of
the analog signal $y_m(t)$. To approximate the PAPR of the analog
signal accurately, an $L$-times oversampled version of the analog
signal is usually considered \cite{Tellambura01}, that is
\begin{align}
y_{mk}=\frac{1}{\sqrt{N}}\sum_{n=0}^{N-1}x_{mn}\cdot e^{\frac{j2\pi nk}{LN}},~ 0\leq k\leq LN-1
\label{IDFT}
\end{align}
where $L$ is an integer no less than $1$. This oversampling
operation (\ref{IDFT}) can also be expressed as
\begin{align}
\boldsymbol{y}_m=\boldsymbol{F}_{LN}^H\boldsymbol{x}_m
\end{align}
where the oversampling matrix
$\boldsymbol{F}_{LN}^H\in\mathbb{C}^{LN\times N}$ is the first $N$
columns of the $LN$-points IDFT matrix (scaled by $\sqrt{L}$).
Thus the PAPR of the time-domain samples with $L$-times
oversampling is given by
\begin{align}
\textsf{PAPR}\left(\boldsymbol{y}_m\right)
=\frac{\max\limits_{0\leq{k}\leq{LN-1}}|y_{mk}|^2}
{\mathbb{E}\{|y_{mk}|^2\}}
=\frac{LN\left\|\boldsymbol{y}_m\right\|_\infty^2}{\left\|\boldsymbol{y}_m\right\|_2^2}
\label{PAPR}
\end{align}
For any signals $\boldsymbol{y}_m\in\mathbb{C}^{LN\times1}$, the
PAPR satisfies the following inequalities:
\begin{align}
1\leq\textsf{PAPR}\left(\boldsymbol{y}_m\right)\leq LN
\label{PAPRbound}
\end{align}
So far most existing PAPR reduction methods (e.g.
\cite{StuderLarsson13,BaoFang2016,ChenWang15}) for large-scale MU
MIMO-OFDM systems reduce the PAPR through directly devising the
precoded signals $\{\boldsymbol{w}_n\}$. The rationale behind
these works is that, due to the redundant degrees-of-freedom
(DoFs) rendered by the large number of antennas at the BS, there
exist an infinite number of precoded signals that can achieve
perfect MUI cancelation, from which we may find a set of precoded
signals $\{\boldsymbol{w}_n\}$ whose time-domain counterpart
signals $\{\boldsymbol{y}_m\}$ have a low PAPR. These
precoding-based approaches, however, may have limited
applicability because, in practical systems, the precoding
matrices have to be chosen from a pre-specified codebook, and as a
result, the precoded signals cannot be arbitrarily devised. To
address this difficulty, in the following, we propose a
perturbation-assisted scheme which does not rely on the precoding
design to reduce the PAPR.

\section{Proposed Perturbation-Assisted PAPR Reduction Method}
The idea of our PAPR reduction method is to add carefully designed
perturbation signals $\{\Delta\boldsymbol{x}_m\}$ to the precoded
signals $\{\boldsymbol{x}_m\}$ (a reordered version of
$\{\boldsymbol{w}_n\}$) to reduce the PAPR of the resulting
time-domain signals $\{\boldsymbol{y}_m\}$ (see Fig.
\ref{systemmodel}). Meanwhile, the additive perturbation signal is
constrained to lie in the null space of its associate channel
matrix such that it is invisible to the receivers, i.e. the signal
received by $K$ users remains unchanged before and after the
perturbation signals are added to the precoded signals. We assume
that a zero-forcing precoding or other precoding schemes such as a
dirty paper coding is employed to remove or suppress the
multi-user interference. Since the perturbation signals vanish
after propagating through the wireless channel, inclusion of the
perturbation signals does not incur additional multi-user
interference. Therefore, unlike previous works (e.g.
\cite{StuderLarsson13,BaoFang2016,ChenWang15}) that jointly
consider the PAPR reduction and multi-user precoding, the PAPR
reduction problem is decoupled from the multi-user precoding in
our paper. As will be shown in this paper, this decoupling enables
us to develop a more efficient algorithm that has a faster
convergence rate than existing methods.


Let
\begin{align}
\Delta\boldsymbol{X}\triangleq[\Delta\boldsymbol{x}_1\cdot\cdot\cdot\Delta\boldsymbol{x}_M]
=[(\Delta\boldsymbol{x}^r_1)^T\cdot\cdot\cdot(\Delta\boldsymbol{x}^r_N)^T]^T
\end{align}
denote the perturbation signals added to the precoded signals
$\{\boldsymbol{x}_m\}$. Specifically,
$\Delta\boldsymbol{x}_m\in\mathbb{C}^{N\times1}$ is the
perturbation signal added to the precoded signal of the $m$-th
transmit antenna, i.e. $\boldsymbol{x}_m$, while
$\Delta\boldsymbol{x}^r_n\in\mathbb{C}^{1\times M}$ (the $n$th row
of $\Delta\boldsymbol{X}$) is the perturbation signal added to the
$n$-th tone signal, i.e. $\boldsymbol{w}_n$. To avoid any in-band
distortions in the frequency domain, the perturbation signals are
required to satisfy the following conditions:
\begin{align}
\boldsymbol{H}_n(\Delta\boldsymbol{x}^r_n)^T = \boldsymbol{0}_{K\times1},\quad  n\in\mathcal{T}
\label{con1}
\end{align}
That is, the perturbation signal added to the $n$th tone has to
lie in the null space of the channel matrix associated with the
$n$th tone. This constraint guarantees that the perturbation
signals do not cause any multi-user inference to receivers. Also,
to avoid out-of-band radiations, we impose the following
constraints on the guard bands:
\begin{align}
\Delta\boldsymbol{x}^r_n = \boldsymbol{0}_{1\times M},  \quad n\in\mathcal{T}^\mathcal{C}
\label{con2}
\end{align}
With the above two constraints, the addition of the perturbation
signal $\Delta\boldsymbol{X}$ does not necessitate any additional
processing at the receivers since the perturbation signals are
like nonexistent to the receivers.



We now discuss how to design the perturbation signals
$\{\Delta\boldsymbol{x}_m\}$ to reduce the PAPR at each transmit
antenna. Ideally, we wish to minimize the PAPR (\ref{PAPR})
associated with each transmit antenna. This, however, is an
ill-defined problem because the above constraints
(\ref{con1})--(\ref{con2}) demands a joint optimization of the
perturbations signals $\{\Delta\boldsymbol{x}_m\}$, and hence
PAPRs of different antennas cannot be simultaneously minimized.
This is also the situation for other works, e.g.
\cite{StuderLarsson13,BaoFang2016,ChenWang15}. Besides, directly
optimizing (\ref{PAPR}) results in a non-convex problem and hence,
finding the solution with an efficient algorithm seems to be
difficult. To circumvent this difficulty, \cite{StuderLarsson13}
proposes to minimize the largest magnitude of the time-domain
signal samples, which not only leads to a convex formulation, but
also has been shown to be effective to substantially reduce the
PAPR. Inspired by \cite{StuderLarsson13}, we propose to minimize a
sum of the largest magnitudes of different antenna's time-domain
signals. The problem can be cast as
\begin{equation}
\begin{split}
&\mathop{\text{minimize}}\limits_{\Delta\boldsymbol{X}} \quad \sum_{m=1}^M\left\|\boldsymbol{y}_m\right\|_\infty\label{obj}\\
&\text{subject to}\quad
\left\{
\begin{aligned}
&\boldsymbol{Y} = \boldsymbol{F}_{LN}^H(\boldsymbol{X}+\Delta\boldsymbol{X}) \\
&\boldsymbol{H}_n(\Delta\boldsymbol{x}^r_n)^T = \boldsymbol{0}_{K\times1},\quad  n\in\mathcal{T} \\
&\Delta\boldsymbol{x}^r_n = \boldsymbol{0}_{1\times M},  \quad n\in\mathcal{T}^\mathcal{C}
\end{aligned}
\right.
\end{split}
\end{equation}
where $\ell_{\infty}$ denotes the infinity-norm,
$\boldsymbol{X}\triangleq
[\boldsymbol{x}_1\phantom{0}\ldots\phantom{0}\boldsymbol{x}_M]$ is
the aggregation of all antenna's precoded frequency-domain
signals, and
$\boldsymbol{Y}\triangleq[\boldsymbol{y}_{1}\cdot\cdot\cdot\boldsymbol{y}_{M}]$
denotes the time-domain signal associated with $M$ transmit
antennas. Note that unlike the method \cite{StuderLarsson13} that
minimizes the largest magnitude of all antennas' signals (i.e.
$\|\text{vec}(\boldsymbol{Y})\|_\infty$), in our proposed scheme,
the sum of each antenna's largest magnitudes is minimized. This
metric is useful in the sense that one can assign different
weights to different $\ell_\infty$-norm terms, which may be needed
when multiple types of RF chains are installed at the BS.





The inclusion of the perturbation signal may result in an increase
of the transmit power. To address this issue, we can impose a
pre-specified upper bound $P_\text{max}$ on the transmit power,
i.e. $\|\boldsymbol{X}+\Delta\boldsymbol{X}\|_F^2\leq
P_\text{max}$. If a ZF precoding scheme is employed, we have
\begin{align}
\big\|\boldsymbol{X}+\Delta\boldsymbol{{X}}\big\|_F^2
&=\big\|\boldsymbol{X}\big\|_F^2+\big\|\Delta\boldsymbol{{X}}\big\|_F^2  \nonumber\\
&~~+\sum_{n=1}^N\left[(\Delta\boldsymbol{x}^r_n)^*{\boldsymbol{w}}_n
+\boldsymbol{w}_{n}^H(\Delta\boldsymbol{{x}}^r_n)^T\right]\nonumber\\
&=\big\|\boldsymbol{X}\big\|_F^2+\big\|\Delta\boldsymbol{{X}}\big\|_F^2
\label{powerequation}
\end{align}
where the second equality follows from the constraint
(\ref{con1}), i.e.
\begin{align}
(\Delta\boldsymbol{{x}}^r_n)^* \boldsymbol{w}_n
&=(\Delta\boldsymbol{{x}}^r_n)^* \boldsymbol{H}_n^H(\boldsymbol{H}_n\boldsymbol{H}_n^H)^{-1}\boldsymbol{s}_n\nonumber\\
&=\boldsymbol{0}_{1\times K}(\boldsymbol{H}_n\boldsymbol{H}_n^H)^{-1}\boldsymbol{s}_n
=0,\quad n\in \mathcal{T}
\end{align}
Equation (\ref{powerequation}) indicates that, if a ZF precoding
scheme is employed, adding a perturbation signal to the precoded
signal always results in an increase in the transmit power, and
the increment is exactly the amount of power of the perturbation
signal. In fact, for other linear precoding schemes such as the MF
precoding and the MMSE precoding, it can be easily verified that
(\ref{powerequation}) holds valid as well. Therefore, we can
simply impose a power constraint on the perturbation signal, i.e.
$\|\Delta\boldsymbol{X}\|_F^2\leq \Delta P_\text{max}$. Note that
including this constraint into (\ref{obj}) does not change the
convexity of the optimization problem. Nevertheless, to facilitate
an efficient algorithm development, this power constraint is
omitted in this paper. On the other hand, our simulation results
suggests that this power constraint may not be needed since as
illustrated in our simulations, the power increase is always small
even without taking this power constraint into account.

\section{PROXINF-ADMM Algorithm}
In this section, we develop an efficient PAPR reduction algorithm
to find an effective solution to (\ref{obj}). To make the problem
(\ref{obj}) tractable, a variable splitting technique is used,
where $\boldsymbol{Y}$ in (\ref{obj}) are treated as splitting
variables and the equality constraint $\boldsymbol{Y} =
\boldsymbol{F}_{LN}^H(\boldsymbol{X}+\Delta\boldsymbol{X})$ is
relaxed as a Lagrange multiplier
\begin{align}
\begin{split}
&\mathop{\text{minimize}}\limits_{\boldsymbol{Y},
\Delta\boldsymbol{X}} ~~\lambda\sum_{m=1}^M\big\|\boldsymbol{y}_m\big\|_\infty+
\big\|\boldsymbol{Y}\!-\!\boldsymbol{F}_{LN}^H(\boldsymbol{X}\!+\!\Delta\boldsymbol{X})\big\|_F^2\\
&\text{subject to}~~
\left\{
\begin{aligned}
&\boldsymbol{H}_n(\Delta\boldsymbol{x}^r_n)^T = \boldsymbol{0}_{K\times1},\quad  n\in\mathcal{T} \\
&\Delta\boldsymbol{x}^r_n = \boldsymbol{0}_{1\times M}, \quad n\in\mathcal{T}^\mathcal{C}
\end{aligned}
\right.
\end{split}
\label{relax-obj}
\end{align}
where $\lambda>0$ is a regularization parameter whose choice will
be elaborated later. Note that the signal to be transmitted is
$\boldsymbol{F}_{LN}^H(\boldsymbol{X}\!+\!\Delta\boldsymbol{X})$,
instead of $\boldsymbol{Y}$. Hence, this relaxation does not cause
any additional distortions or multi-user interference as long as
$\Delta\boldsymbol{X}$ satisfies the constraints in
(\ref{relax-obj}). The variable splitting allows the original
intractable optimization problem to be decomposed into tractable
sub-problems. Specifically, an alternating minimization strategy
can be used to solve (\ref{relax-obj}), in which we alternatively
minimize the objective function with respect to $\boldsymbol{Y}$
and $\Delta\boldsymbol{X}$. This alternating minimization ensures
that the objective function is non-increasing at each iteration.

We now discuss how to alternatively minimize the objective
function in (\ref{relax-obj}) with respect to $\boldsymbol{Y}$ and
$\Delta\boldsymbol{X}$. To facilitate our exposition, the
objective function in (\ref{relax-obj}) is denoted as
$f(\Delta\boldsymbol{X},\boldsymbol{Y})$, and we use $\mathcal{F}$
to denote the feasible set of $\Delta\boldsymbol{X}$. Thus in the
$(t+1)$th iteration, the alternating procedure can be expressed as
\begin{align}
\boldsymbol{Y}^{(t+1)}=&\mathop{\text{argmin}}\limits_{\boldsymbol{Y}}
f(\Delta\boldsymbol{X}^{(t)},\boldsymbol{Y})\label{Y-update}\\
\Delta\boldsymbol{X}^{(t+1)}=&\mathop{\text{argmin}}\limits_{\Delta\boldsymbol{X}}
f(\Delta\boldsymbol{X},\boldsymbol{Y}^{(t+1)}),~~\Delta\boldsymbol{X}\in\mathcal{F}
\label{X-update}
\end{align}
Let us first consider the optimization of $\boldsymbol{Y}$.
Clearly, the optimization (\ref{Y-update}) can be decomposed into
$M$ independent subproblems, each of which is known as the
proximal operator of the $\ell_\infty$-norm
\cite{parikh2013proximal,studer2014democratic}
\begin{align}
\textsf{PROXINF}(\boldsymbol{q}_m^{(t)},\lambda)=\mathop{\text{argmin}}\limits_{\boldsymbol{y}_m}
\lambda\big\|\boldsymbol{y}_m\big\|_\infty\!+\!\big\|\boldsymbol{y}_m\!-\!\boldsymbol{q}_m^{(t)}\big\|_2^2
\label{proxinf}
\end{align}
where
\begin{align}
\boldsymbol{q}_m^{(t)}\triangleq\boldsymbol{F}_{LN}^H(\boldsymbol{x}_m+\Delta\boldsymbol{x}_m^{(t)})
\end{align}
As shown in \cite{StuderLarsson13}, the proximal operator of the
$\ell_\infty$-norm is in fact a clipping operator, in which the
vector $\boldsymbol{q}_m^{(t)}$ is clipped by a clipping level $A$
that is controlled by the regularization parameter $\lambda$. Thus
in the step of $\boldsymbol{Y}$-update, the proposed algorithm
clips the peaks of the transmit signal associated with each
transmit antenna. The clipping level $A$ does not have a
closed-form solution. Nevertheless, the value of $A$ can be
efficiently obtained by resorting to the method (i.e. Algorithm 2)
developed in \cite{studer2014democratic}.

Concerning the update of ${\Delta\boldsymbol{X}}$, the
optimization of ${\Delta\boldsymbol{X}}$ can be expressed as
\begin{align}
\mathop{\text{minimize}}\limits_{\Delta\boldsymbol{X}}~~
\big\|\boldsymbol{V}^{(t+1)}-\boldsymbol{F}_{LN}^H\Delta\boldsymbol{X}\big\|_F^2,~~\Delta\boldsymbol{X}\in\mathcal{F}
\label{X-update1}
\end{align}
where
\begin{align}
\boldsymbol{V}^{(t+1)}\triangleq\boldsymbol{Y}^{(t+1)}-\boldsymbol{F}_{LN}^H\boldsymbol{X}
\end{align}
The minimization of $\Delta\boldsymbol{X}$ is not straightforward
due to the constraints that $\Delta\boldsymbol{X}$ has to satisfy.
In the following, we resort to the alternating direction method of
multipliers (ADMM) technique to solve (\ref{X-update1}).




\subsection{Solve (\ref{X-update1}) via ADMM}
The alternating direction method of multipliers (ADMM) \cite{ADMM}
is a simple but powerful technique that solves convex optimization
problems by breaking them into smaller pieces, each of which is
then easier to handle. First, we use $\boldsymbol{D}$ to denote
the optimization variable $\Delta\boldsymbol{X}$. The optimization
(\ref{X-update1}) can be rewritten as
\begin{align}
\mathop{\text{minimize}}\limits_{\boldsymbol{D}}~~
\big\|\boldsymbol{V}^{(t+1)}-\boldsymbol{F}_{LN}^H\boldsymbol{D}\big\|_F^2,~~\boldsymbol{D}\in\mathcal{F}
\label{ADMM1}
\end{align}
By introducing an auxiliary variable $\boldsymbol{Z}$, the above
optimization (\ref{ADMM1}) can be equivalently written as
\begin{equation}
\begin{split}
&\mathop{\text{minimize}}\limits_{\boldsymbol{D},\boldsymbol{Z}}~~
\big\|\boldsymbol{V}^{(t+1)}-\boldsymbol{F}_{LN}^H\boldsymbol{Z}\big\|_F^2\\
&\text{subject to}~~
\left\{
\begin{aligned}
&\boldsymbol{D}-\boldsymbol{Z}=\boldsymbol{0}_{N\times M}\\
&\boldsymbol{D}\in\mathcal{F}
\end{aligned}
\right.
\end{split} \label{opt-1}
\end{equation}
We resort to the ADMM to solve the above optimization
(\ref{opt-1}). We first form an \textit{augmented Lagrangian} of
the above optimization problem
\begin{align}
\mathcal{L}(\boldsymbol{Z},&\boldsymbol{D},\boldsymbol{U})\nonumber\\
=&\big\|\boldsymbol{V}^{(t+1)}-\boldsymbol{F}_{LN}^H\boldsymbol{Z}\big\|_F^2
+\rho\big\|\boldsymbol{D}-\boldsymbol{Z}\big\|_F^2\\
&+\sum_{n=1}^N\sum_{m=1}^M
\left[u_{nm}^*(d_{nm}-z_{nm})+u_{nm}(d_{nm}-z_{nm})^*\right]\nonumber
\end{align}
where $\rho>0$ is called the penalty parameter, $d_{nm}$ and
$z_{nm}$ denote the $(n,m)$th entry of $\boldsymbol{D}$ and
$\boldsymbol{Z}$, respectively,
$\boldsymbol{U}\in\mathbb{C}^{N\times M}$ is the dual variable or
Lagrange multiplier, and $u_{nm}$ denotes the $(n,m)$th entry of
$\boldsymbol{U}$. The optimization variables $\boldsymbol{Z}$ and
$\boldsymbol{D}$, along with the dual variable $\boldsymbol{U}$,
can be optimized via the following iterations \cite{ADMM}:
\begin{align}
\boldsymbol{Z}^{(i+1)}&=\mathop{\text{argmin}}\limits_{\boldsymbol{Z}}
\mathcal{L}(\boldsymbol{Z},\boldsymbol{D}^{(i)},\boldsymbol{U}^{(i)})\label{Z-update}\\
\boldsymbol{D}^{(i+1)}&=\mathop{\text{argmin}}\limits_{\boldsymbol{D}}
\mathcal{L}(\boldsymbol{Z}^{(i+1)},\boldsymbol{D},\boldsymbol{U}^{(i)}),~~\boldsymbol{D}\in\mathcal{F}\label{D-update}\\
\boldsymbol{U}^{(i+1)}&=\boldsymbol{U}^{(i)}+\rho(\boldsymbol{D}^{(i+1)}-\boldsymbol{Z}^{(i+1)})
\end{align}
where the superscript $i$ denotes the number of iterations. We see
that in each iteration, the ADMM algorithm consists of a
$\boldsymbol{Z}$-minimization step, a
$\boldsymbol{D}$-minimization step, and a dual variable update.
For the $\boldsymbol{Z}$-minimization problem, by setting the
derivative of the augmented Lagrangian
$\mathcal{L}(\boldsymbol{Z},\boldsymbol{D}^{(i)},\boldsymbol{U}^{(i)})$
with respect to $\boldsymbol{Z}$ to zero, a closed-form solution
of $\boldsymbol{Z}^{(i+1)}$ can be obtained as
\begin{align}
\boldsymbol{Z}^{(i+1)}=
\frac{(\boldsymbol{A}^{(t+1)}+\rho\boldsymbol{D}^{(i)}+\boldsymbol{U}^{(i)})}{(1+\rho)}
\end{align}
where
$\boldsymbol{A}^{(t+1)}\triangleq\boldsymbol{F}_{LN}\boldsymbol{V}^{(t+1)}$.
For the $\boldsymbol{D}$-minimization problem (\ref{D-update}),
after some algebraic manipulations, it can be further decomposed
into $|\mathcal{T}|$ independent subproblems, i.e.
\begin{align}
\begin{split}
&\mathop{\text{minimize}}\limits_{\boldsymbol{d}_n^r}\quad
\big\|\boldsymbol{d}_n^r-(\boldsymbol{z}_n^{r\:(i+1)}-
\boldsymbol{u}_n^{r\:(i)}/\rho)\big\|_F^2,~ \forall n\in\mathcal{T}\\
&\text{subject to}\quad \boldsymbol{H}_n(\boldsymbol{d}_n^{r})^T=\boldsymbol{0}_{K\times1}
\end{split} \label{opt-2}
\end{align}
where $\boldsymbol{d}_n^r$, $\boldsymbol{z}_n^r$, and
$\boldsymbol{u}_n^r$ denote the $n$th row of $\boldsymbol{D}$,
$\boldsymbol{Z}$, and $\boldsymbol{U}$, respectively. For each
tone $n\in\mathcal{T}$, the optimization (\ref{opt-2}) searches
for a vector that is nearest to the vector
$(\boldsymbol{z}_n^{r\:(i+1)}-\boldsymbol{u}_n^{r\:(i)}/\rho)$,
and meanwhile lies in the null space of the channel matrix
$\boldsymbol{H}_n$. The optimal solution, clearly, is the
projection of the vector
$(\boldsymbol{z}_n^{r\:(i+1)}-\boldsymbol{u}_n^{r\:(i)}/\rho)$
onto the null space of $\boldsymbol{H}_n$, which is given by
\begin{align}
\boldsymbol{d}_n^{r\:(i+1)}=(\boldsymbol{z}_n^{r\:(i+1)}-\boldsymbol{u}_n^{r\:(i)}/\rho)\boldsymbol{G}_n^T
\end{align}
where $\boldsymbol{G}_n$ denotes the orthogonal projection onto
the null-space of $\boldsymbol{H}_n$, i.e.
\begin{align}
\boldsymbol{G}_n=
\boldsymbol{I}_M-\boldsymbol{H}_n^H(\boldsymbol{H}_n\boldsymbol{H}_n^H)^{-1}\boldsymbol{H}_n
\end{align}
Note that for $\forall n\in\mathcal{T}^\mathcal{C}$, we always have $\boldsymbol{d}_n^{r}=\boldsymbol{0}_{1\times M}$.

\subsection{Summary}
The proposed algorithm is referred to as the PROXINF-ADMM
algorithm which proceeds in a double-loop manner: the outer loop
clips the peaks of the transmitted time-domain signals
$\{\boldsymbol{q}_m^{(t)}\}$ via the $\ell_{\infty}$-norm proximal
operator, and the inner loop updates the perturbation signal via
the ADMM algorithm. The details of the PROXINF-ADMM algorithm is
summarized in the following table. Our simulation results suggest
that only very few iterations are needed to implement the inner
loop, i.e. there is no need to wait until the ADMM algorithm
converges, and this early termination of the inner loop does not
affect the convergence of the proposed algorithm. The dominating
operations in each iteration is the simple matrix-vector
multiplications, which scale as $\mathcal{O}(MN)$. Thus the
proposed algorithm has a low computational complexity. Also, note
that the constraints (\ref{con1}) and (\ref{con2}) are always
satisfied throughout the while iterative process. Hence any
intermediate solution $\Delta\boldsymbol{X}^{(t)}$ can be used,
without causing any in-band or out-of-band radiations. This merit
is useful in practical systems.

\begin{table}[t]
\normalsize
\begin{center}
\textbf{PROXINF-ADMM Algorithm}
\end{center}
\doublerulesep=0.4pt \noindent
\begin{tabular}{p{8.45cm}}

\hline\hline \vskip -0.2 cm Given a frequency-domain signal
$\boldsymbol{X}\!\in\!\mathbb{C}^{N\times M}$, devise a
perturbation signal
$\Delta\boldsymbol{X}\!\in\!\mathbb{C}^{N\times M}$.
\begin{enumerate}[1)]
\item \textit{Initialization:} Set $t=0$, $\Delta\boldsymbol{X}^{(0)}=\boldsymbol{0}_{N\times M}$,
and set $\lambda$ and $\rho$ to some initial values, and compute
\begin{equation}
\begin{aligned}
~~~~~~~~~\boldsymbol{G}_n=
\boldsymbol{I}_M\!-\!\boldsymbol{H}_n^H(\boldsymbol{H}_n\boldsymbol{H}_n^H)^{-1}
\!\boldsymbol{H}_n,\:\forall n\in \mathcal{T}\nonumber
\end{aligned}
\end{equation}


\item \textit{Outer loop, update $\boldsymbol{Y}$:}
For $m=1,...,M$, do
\begin{equation}
\begin{aligned}
\boldsymbol{q}_m^{(t)}=\boldsymbol{F}_{LN}^H(\boldsymbol{x}_m+\Delta\boldsymbol{x}_m^{(t)})~~~~~~~~~~\nonumber
\end{aligned}
\end{equation}
\begin{equation}
\begin{aligned}
\boldsymbol{y}_m^{(t+1)}=\textsf{PROXINF}(\boldsymbol{q}_m^{(t)},\lambda)~~~~~~~~~~~~~~\nonumber
\end{aligned}
\end{equation}

\item \textit{ADMM inner loop:} Set
$\boldsymbol{D}^{(0)}=\Delta\boldsymbol{X}^{(t)}$ and
$\boldsymbol{U}^{(0)}=\boldsymbol{0}_{N\times M}$, and compute
\begin{align}
\boldsymbol{A}^{(t+1)}=\boldsymbol{F}_{LN}\boldsymbol{Y}^{(t+1)}-\boldsymbol{X}~~~~~~~~~~~~~~~\,\nonumber
\end{align}
and for $i\!=\!0,1,...,I_\text{max}\!-\!1$, repeat the following recursions
\begin{equation}
\begin{aligned}
\boldsymbol{Z}^{(i+1)}=
\frac{(\boldsymbol{A}^{(t+1)}+\rho\boldsymbol{D}^{(i)}+\boldsymbol{U}^{(i)})}{(1+\rho)}~~~~~~\nonumber
\end{aligned}
\end{equation}
\begin{equation}
\begin{aligned}
~~\boldsymbol{d}_n^{r\:(i+1)}=(\boldsymbol{z}_n^{r\:(i+1)}-\boldsymbol{u}_n^{r\:(i)}/\rho)\boldsymbol{G}_n^T
,~\forall n\in \mathcal{T}\nonumber
\end{aligned}
\end{equation}
\begin{equation}
\begin{aligned}
\boldsymbol{U}^{(i+1)}=\boldsymbol{U}^{(i)}+\rho(\boldsymbol{D}^{(i+1)}-\boldsymbol{Z}^{(i+1)})~~~~
\nonumber
\end{aligned}
\end{equation}

\item Set $\Delta\boldsymbol{X}^{(t+1)}\!=\!\boldsymbol{D}^{(I_\text{max})}$ and increase $t=t+1$,
return to step 2 if $t<T_\text{max}$, otherwise stop and return the solution.
\end{enumerate}
\begin{spacing}{-1}
\end{spacing}\\
\hline\hline
\end{tabular}
\end{table}

\section{Simulation Results}
In this section, we carry out experiments to illustrate the
performance of the proposed PAPR reduction
algorithm\footnote{Codes are available at
http://www.junfang-uestc.net/codes/PROXINF-ADMM.rar} (referred to
as the PROXINF-ADMM). We compare our method with the FITRA
algorithm \cite{StuderLarsson13}, the zero-forcing (ZF) precoding
scheme, and the amplitude clipping scheme.

In our simulations, the BS is assumed to have $M=128$ transmit
antennas and serve $K=16$ single-antenna users. We consider an
OFDM modulation with $N=128$ tones and use a spectral map
$\mathcal{T}$ as specified in the 40 MHz mode of Wi-Fi
\cite{Wi-Fi}, in which $|\mathcal{T}|=114$ tones are used for data
transmission. We also consider the convolutional-coded
transmission where the information bits for each user are first
encoded by a convolutional encoder with generator polynomials
$[5_o~7_o]$, and then are randomly interleaved and mapped to a
64-QAM constellation (Gray-coded). The wireless channel is assumed
be frequency-selective and modeled as a tap-delay line with $D=8$
taps. The time-domain channel response matrices
$\boldsymbol{\hat{H}}_{d}$, $d=1,...,D$, have i.i.d. circularly
symmetric Gaussian distributed entries with zero mean and unit
variance, and the equivalent frequency-domain response
$\boldsymbol{H}_n$ on the $n$-th tone can be obtained by
\begin{align}
\boldsymbol{H}_n=\sum^{D}_{d=1}\boldsymbol{\hat {H}}_{d}\exp\left(\frac{-j2\pi
dn}{N}\right).
\end{align}
In each user terminal, after demodulating the received symbols, a
Viterbi decoder is employed to decode the information bits.

For the PROXINF-ADMM algorithm and the amplitude clipping method,
we assume the precoded signal $\boldsymbol{X}$ is generated by a
ZF precoding scheme. The maximum number of iterations for the
FITRA algorithm is set to be $2000$, as suggested by
\cite{StuderLarsson13}. For our proposed algorithm, unless
explicitly stated otherwise, the maximum numbers of iterations for
the outer loop and the inner loop are set to be $T_\text{max}=200$
and $I_\text{max}=2$, respectively. The regularization parameter
and the penalty parameter are chosen to be $\lambda=1$ and
$\rho=0.5$, respectively. In our simulations, we employ an
oversampling rate of $L=4$, and use the complementary cumulative
distribution function (CCDF) to evaluate the PAPR reduction
performance. The CCDF denotes the probability that the PAPR of the
estimated signal exceeds a given threshold $\textsf{PAPR}_0$, that
is
\begin{align}
\textsf{CCDF}(\textsf{PAPR}_0)=\text{Pr}(\textsf{PAPR}>\textsf{PAPR}_0).
\end{align}
Also, in order to evaluate the increase of the transmit power, we
define the power increase (PI) as
\begin{align}
\textsf{PI}=\frac{\big\|\boldsymbol{\hat{X}}\big\|_F^2}
{\big\|\boldsymbol{X}^\text{ZF}\big\|_F^2},
\end{align}
where $\boldsymbol{\hat{X}}$ denotes the low-PAPR solution
rendered by different schemes, and $\boldsymbol{X}^\text{ZF}$
represents the solution obtained by using a ZF precoding scheme.
Note that for the FITRA and PROXINF-ADMM, we have
$\textsf{PI}>0\,$dB in general, while for the clipping scheme, we
have $\textsf{PI}<0\,$dB. We also note that our proposed method
yields solutions that strictly satisfy the null space constraints.
Therefore it does not cause any multi-user interference (MUI) or
out-of-band radiations. In contrast, the FITRA algorithm cannot
achieve perfect MUI and out-of-band radiation cancelation, and
needs to choose an appropriate regularization parameter to ensure
small MUI and out-of-band radiations.

\begin{figure*}[!t]
\setlength{\abovecaptionskip}{0pt}
\setlength{\belowcaptionskip}{0pt} \centering
\includegraphics[width=17.2cm]{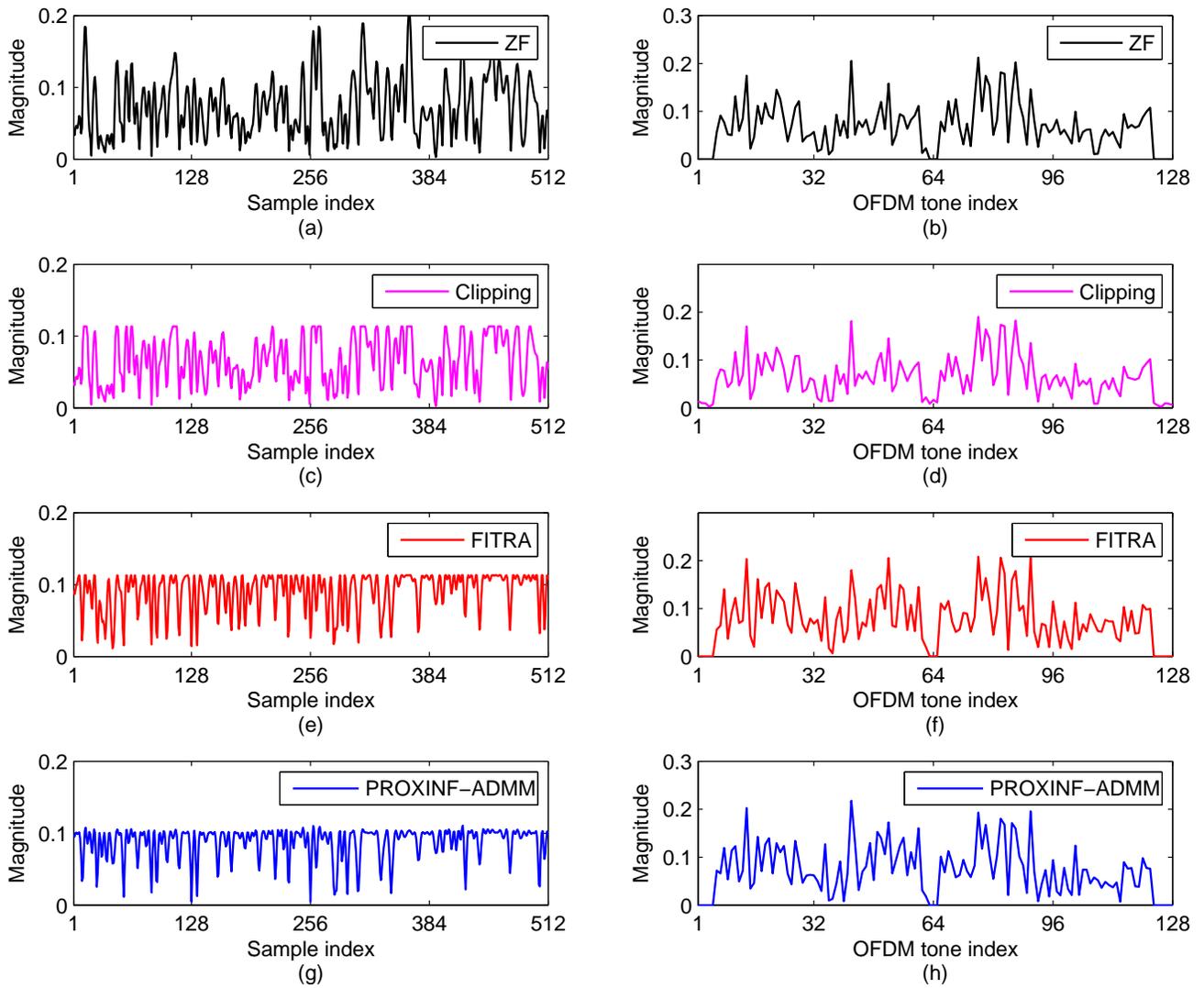}
\caption{Time/frequency representation for different schemes. (a),
(c), (e) and (g) are time-domain signals for ZF, clipping, FITRA
and PROXINF-ADMM, respectively (\textsf{PAPR}:
$\text{ZF}=9.1\,$dB, $\text{Clipping}=3.9\,$dB,
$\text{FITRA}=2.0\,$dB, and $\text{PROXINF-ADMM}=2.2\,$dB). (b),
(d), (f) and (h) are corresponding frequency-domain signals for
respective schemes. } \label{signal}
\end{figure*}

Firstly, we examine the time-domain and frequency-domain signals
obtained by respective schemes. The (a), (c), (e) and (g) of Fig.
\ref{signal} depict the magnitudes of the first antenna's
time-domain samples (i.e. $\boldsymbol{y}_1$) obtained by
respective schemes. We observe that, similar to the FITRA
algorithm, our proposed algorithm yields a quasi-constant
magnitude solution with many of its entries located close to a
ceiling, which leads to a very low PAPR. The solution of the
clipping scheme is only a slightly alleviated version of the ZF
solution. Simulation results show that our proposed method
achieves a PAPR of $2.2\,$dB (PAPR associated with the first
transmit antenna), the FITRA algorithm attains a slightly lower
PAPR of $2.0\,$dB, and the clipping scheme has a higher PAPR of
$3.9\,$dB, while the ZF precoding has the highest PAPR of
$9.1\,$dB. In the (b), (d), (f) and (h) of Fig. \ref{signal}, we
depict the magnitudes of the corresponding frequency-domain
signals. As shown in the figures, there is no out-of-band
radiation for the solutions rendered by the ZF, FITRA and
PROXINF-ADMM methods (the radiation of the FITRA here is
negligible), while the clipping scheme causes severe radiations in
the guard band that could degrade the spectral efficiency
severely.

\begin{figure*}[!t]
\setlength{\abovecaptionskip}{0pt}
\setlength{\belowcaptionskip}{0pt}
\centering
\includegraphics[width=17.5cm]{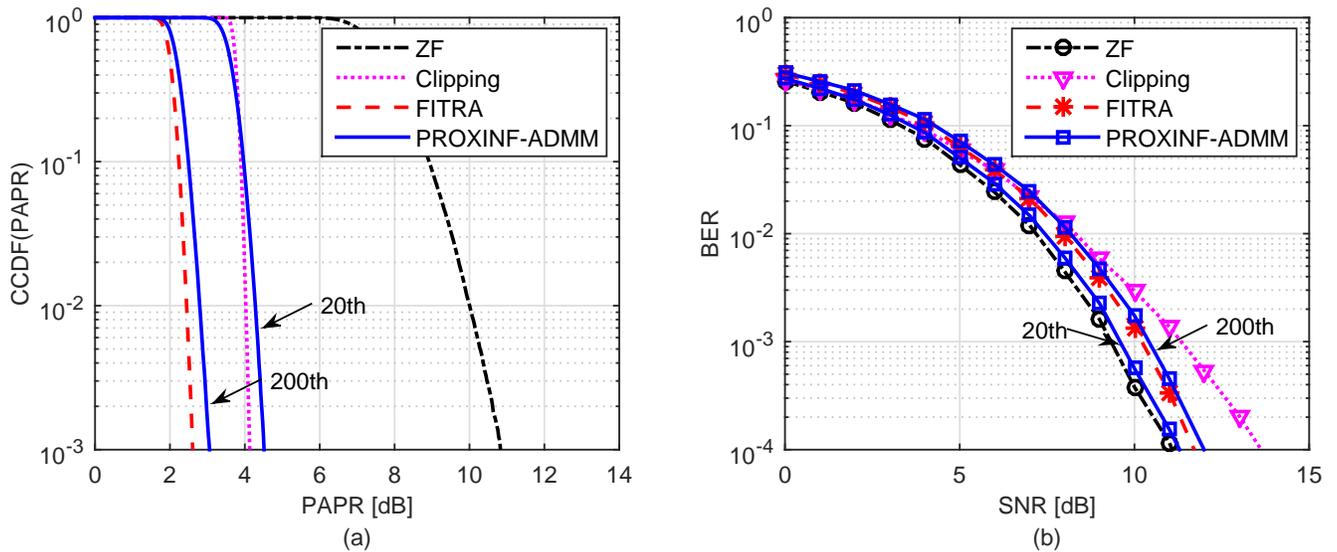}
\caption{CCDFs and bit error rates (BERs) of respective schemes.
(a) CCDFs of respective schemes, (b) BER vs. SNR.}
\label{PAPR_BER}
\end{figure*}

To better evaluate the PAPR reduction performance, we plot the
empirical CCDF of the PAPR for respective schemes in Fig.
\ref{PAPR_BER}(a). The number of trials is chosen to be $1000$ in
our experiments. The PAPR associated with all $M$ transmit
antennas are taken into account to compute the empirical CCDF. We
also include the results of the PROXINF-ADMM algorithm obtained at
the $20$th (outer) iteration. We see that our proposed algorithm,
within only 200 iterations, is able to achieve PAPR reduction
performance similar to the FITRA algorithm that needs to perform
2000 iterations. Also, our proposed method reduces the PAPR by
more than $7\,$dB compared to the ZF scheme (at
$\textsf{CCDF(PAPR)}=0.01$). The bit error rate (BER) performance
of respective algorithms is shown in Fig. \ref{PAPR_BER}(b), where
the signal-to-noise ratio (SNR) is defined as
$\textsf{SNR}=\mathbb{E}\{\|\boldsymbol{\hat{x}}_n^r\|_2^2\}/N_0$,
where $N_0$ denotes the noise variance at the receivers (c.f.
(\ref{transmition})). We can see that both the FITRA and our
proposed method incur an SNR-performance loss of about $1\,$dB
compared to the ZF scheme (at $\textsf{SER}=10^{-4}$). This
performance loss, as discussed in \cite{StuderLarsson13}, is
primarily due to the transmit power increase, i.e. an increase in
the norm of the obtained solution. The performance loss of the
clipping scheme (about $3\,$dB), however, is mainly caused by the
residual MUI. We also observe that the SNR performance gap can be
reduced if we perform only 20 iterations for our proposed method,
in which case the norm of the resulting solution has a less
significant increase.


\begin{figure}[t]
\setlength{\abovecaptionskip}{0pt}
\setlength{\belowcaptionskip}{0pt} \centering
\includegraphics[width=8.3cm]{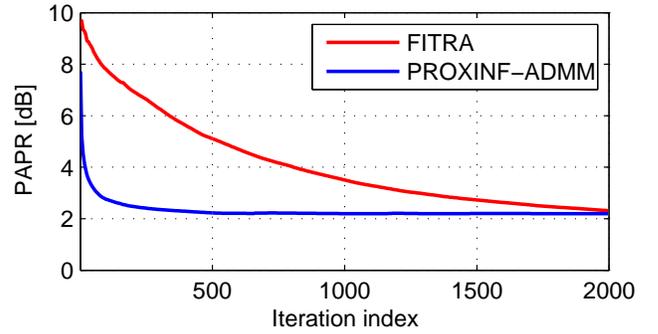}
\caption{PAPRs vs. the number of iterations.} \label{speed}
\end{figure}


We now examine the convergence rates of our proposed method and
the FITRA algorithm. Fig. \ref{speed} shows the PAPR vs. the
number of iterations. We observe that the PROXINF-ADMM algorithm
can obtain a PAPR of $6\,$dB within only several iterations, while
the FITRA algorithm needs about $350$ iterations to reach the same
PAPR reduction performance. Also, the proposed method is able to
reduce the PAPR down to $4\,$dB within only $20$ iterations, while
the FITRA algorithm require as many as $800$ iterations to obtain
a similar result. These results indicate that our proposed method
has a much faster convergence rate than the FITRA algorithm, which
is more suitable for real systems.

\begin{figure*}[!t]
\setlength{\abovecaptionskip}{0pt}
\setlength{\belowcaptionskip}{0pt} \centering
\includegraphics[width=17.5cm]{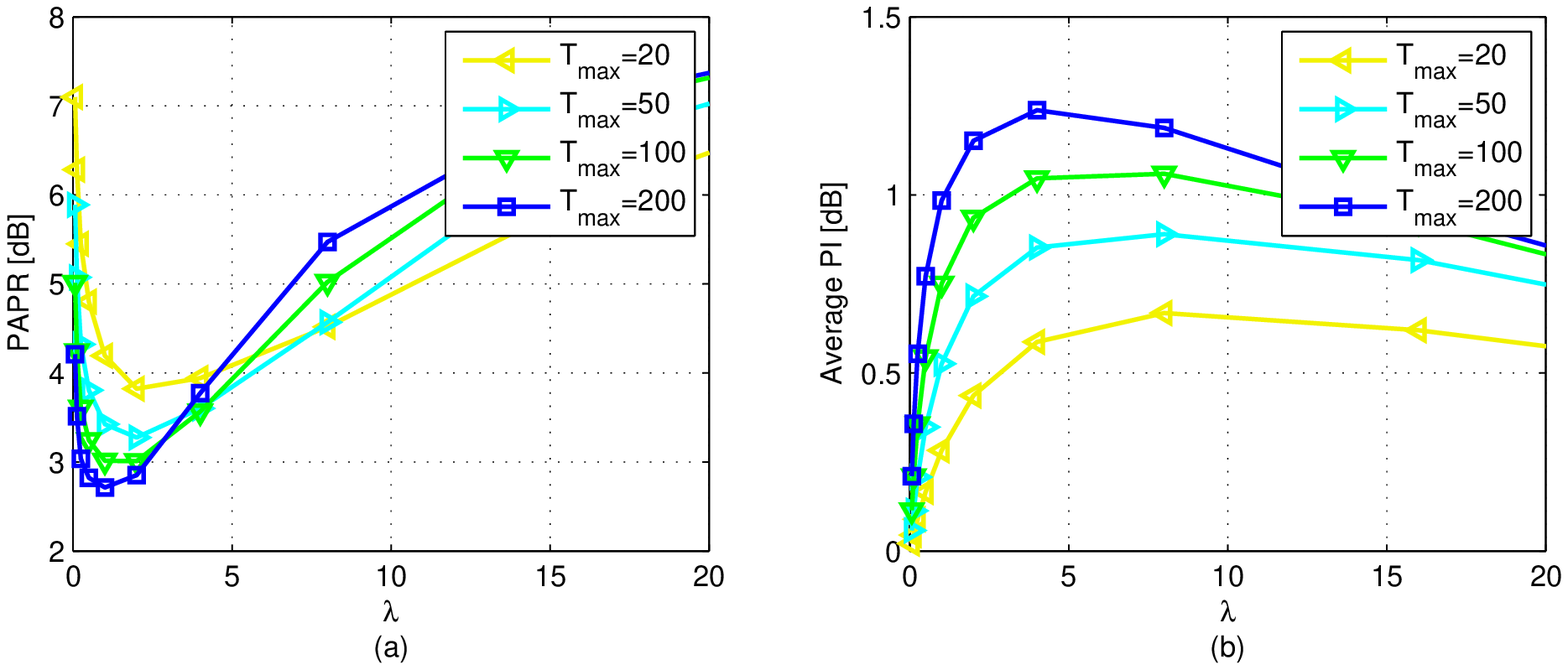}
\caption{PAPR and average PI vs. $\lambda$. (a) PAPR, (b) average
PI.} \label{PAPR_PI_lambda}
\end{figure*}

We examine the impact of the choice of the regularization
parameter $\lambda$ on the PAPR reduction performance. Fig.
\ref{PAPR_PI_lambda} shows the PAPR and the average power increase
(PI) of our proposed method vs. $\lambda$ under different choices
of $T_{\text{max}}$, where $\rho$ is fixed to be $0.5$ and the
results are obtained over $1000$ independent runs. From Fig.
\ref{PAPR_PI_lambda}, we observe that our proposed algorithm is
able to achieve a substantial PAPR reduction when $\lambda$ is
within the range $[0.5,5]$. Moreover, when $\lambda<2$, increasing
the maximum number of iterations $T_\text{max}$ in general reduces
the PAPR but results in a larger power increase (PI). Therefore,
to reduce the PI, one can terminate the iterative process as long
as the solution meets the specified PAPR requirement. Also, an
excessively large value of $\lambda$ leads to bad solutions
because the data fitting term becomes less influential, and as a
result, the transmitted signal
$\boldsymbol{F}_{LN}^H(\boldsymbol{X}\!+\!\Delta\boldsymbol{X})$
could be far away from the desired low PAPR solution
$\boldsymbol{Y}$.

\section{Conclusions}
We considered the problem of PAPR reduction for large-scale
MU-MIMO-OFDM systems. A perturbation-assisted approach was
proposed, where carefully devised artificial perturbation signals
are added to the precoded signals to reduce the PAPRs of the
transmitted signals. Meanwhile, the perturbations signals are
constrained to lie within the null-spaces of the associated
channel matrices such that they cause no multi-user interference
or out-of-band radiations. We formulated the PAPR reduction
problem as a convex optimization problem and developed an
efficient algorithm by resorting to the variable splitting and the
ADMM techniques. Simulations results show that the proposed
algorithm achieves remarkable PAPR reduction performance
comparable to \cite{StuderLarsson13}, meanwhile providing a much
faster convergence rate.


\bibliography{newbib}
\bibliographystyle{IEEEtran}

\end{document}